\documentclass[twocolumn,showpacs,superscriptaddress,preprintnumbers]{revtex4}
\usepackage{graphicx}
\usepackage{dcolumn}
\usepackage{bm}

\begin{document}

\title{Spontaneous Crystallization of Skyrmions and Fractional Vortices in
the Fast-rotating and Rapidly-quenched Spin-1 Bose-Einstein Condensates}
\date{May 20, 2011}
\author{S.-W. Su}
\affiliation{Department of Physics, National Tsing Hua University, Hsinchu 30013, Taiwan}
\author{C.-H. Hsueh}
\affiliation{Department of Physics, National Taiwan Normal University, Taipei 11677,
Taiwan}
\author{I.-K. Liu}
\affiliation{Department of Physics, National Changhua University of Education, Changhua
50058, Taiwan}
\author{T.-L. Horng}
\affiliation{Department of Applied Mathematics, Feng Chia University, Taichung 40724,
Taiwan}
\author{Y.-C. Tsai}
\affiliation{Department of Photonics, Feng Chia University, Taichung 40724, Taiwan}
\author{S.-C. Gou}
\affiliation{Department of Physics, National Changhua University of Education, Changhua
50058, Taiwan}
\author{W. M. Liu}
\affiliation{Instuite of Physics, Chinese Science Academy, Beijing 100190, P. R. China}

\begin{abstract}
We investigate the spontaneous generation of crystallized topological
defects via the combining effects of fast rotation and rapid thermal quench
on the spin-1 Bose-Einstein condensates. By solving the stochastic projected
Gross-Pitaevskii equation, we show that, when the system reaches
equilibrium, a hexagonal lattice of skyrmions, and a square lattice of
half-quantized vortices can be formed in a ferromagnetic and
antiferromagnetic spinor BEC, respetively, which can be imaged by using the
polarization-dependent phase-contrast method.
\end{abstract}

\maketitle

\address{$^{1}$Department of Appled Mathematics, Feng Chia University, Taichung 40724,\\
Taiwan\\
$^{2}$Department of Physics, National Changhua University of Education,\\
Changhua 50058, Taiwan\\
$^{3}$Department of Mathematics, National Taiwan University, Taipei 106,\\
Taiwan}


Topological defects are a manifestation of spontaneously broken symmetries
\cite{Volovik2}. Formation and observation of topological defects are one of
the most fundamental and fascinating topics in various aspects of physics,
ranging from condensed matter physics to cosmology. However, owing to the
limitation of energy scales in the earth-bound physics experiments,
topological defects are mostly created and observed in the condensed matter
systems. For example, magnetic domains walls of magnetized material and
string defects in $^{3}$He superfluid phase transitions have been
extensively studied \cite{Volovik}.

Recently, owing to the realization of spinor Bose-Einstein condensate (BEC)
of alkali atoms in optical trap \cite{Stamper-Kurn,Barrett}, the creation of
topological defects in ultracold atomic systems has become possible. A
spinor BEC is fully characterized by the spin degrees of freedom and behaves
as a vector in the spin space. Theoretical studies for the spinor BEC were
pioneered by Ho \cite{Ho}, and independently by Ohmi and Machida \cite{Ohmi}%
. More recently, spinor BECs with $F>1$ \cite{Ueda1,Schmaljohann,Ueda2} or
with long range dipolar interaction \cite{Tsubota,Ueda3} have also been
theoretically investigated. In general, the physical behavior of the spinor
BEC depends crucially on its magnetic properties, such that the interplay
between the superfluidity and magnetism of the condensate makes the spinor
BEC a candidate to exhibit a variety of nontrivial ordered states, such as
the skyrmions \cite{Al Khawaja}, Mermin-Ho vortices \cite{Mermin},
Anderson-Thouless vortices \cite{Anderson} and so forth.

So far, all theoretical studies regarding the formation of topological
defects in spinor BEC appeal to manipulate the external and internal degrees
of freedom of the condensed atoms at zero temperature. On the other hand,
according to the Kibble-Zurek scenario, topological defects can also be
created through phase transitions at finite temperatures, which are
fundamentally caused by spontaneous symmetry breaking and thermal
fluctuations near the critical point. In this paper, we show that it is
possible to create crystalline orders of skyrmions and fractional vortices
simply by thermally quenching a rotating spin-1 BEC. This enables us to
probe into the very fundamental aspects of topological defects without any
engineering of dynamical processes, since evaporative cooling is
prerequisite in creating BECs and the methods of rotating condensates have
been well developed in a variety of ultracold atoms experiments. In the
framework of mean field theory, the dynamics of a BEC at nonzero
temperatures can be described by the stochastic projected Gross-Pitaevskii
equation (SPGPE)\textit{\ }\cite{A. S. Bradley}, which relies on the
assumption that the system can be treated as a condensate band in contact
with a thermal reservoir comprising of all non-condensed particles. In such
a scheme, the condensate band is described by the truncated Wigner method
\cite{Sinatra} including the projected c-field method, while the
non-condensate band is described by the quantum kinetic theory \cite%
{Gardiner1,Gardiner2}. In the following, we shall solve the SPGPE
numerically for a rotating trapped spin-1 BEC. We show that when the system
is quenched down to a very low temperature, a lattice of skyrmions and
half-quantized vortices (HQVs) can be created in the spinor BEC of $^{87}$Rb
and $^{23}$Na, respectively.

The spin-1 BEC is characterized by a vectorial order parameter, $\mathbf{%
\Psi =}\left(
\begin{array}{ccc}
\Psi _{1}, & \Psi _{0}, & \Psi _{-1}%
\end{array}%
\right) ^{T}$ (the superscript $T$ stands for the transpose), where $\Psi
_{j}$ $\left( j=\pm 1,0\right) $ denotes the macroscopic wave function of
the atoms condensed in the spin states, $\left\vert F=1,m_{F}=\pm
1,0\right\rangle $, respectively. The dynamics of $\Psi _{j}$ in a confining
potential is described by the following coupled nonlinear Schr\"{o}dinger
equations
\begin{eqnarray}
&&i\hbar \partial _{t}\Psi _{j}=\hat{H}_{j}^{GP}\Psi _{j}\qquad  \label{GPE}
\\
&=&\mathcal{\hat{H}}\Psi _{j}+g_{s}\sum\limits_{\alpha
=x,y,z}\sum\limits_{n,k,l=0,\pm 1}\left( \hat{F}_{\alpha }\right)
_{jn}\left( \hat{F}_{\alpha }\right) _{kl}\Psi _{n}\Psi _{k}^{\ast }\Psi _{l}
\nonumber
\end{eqnarray}%
where $\mathcal{\hat{H}}=-\hbar ^{2}\bigtriangledown ^{2}/2m+V(\mathbf{r})+$
$g_{n}\left\vert \mathbf{\Psi }\right\vert ^{2}$ denotes the
spin-independent part of the Hamiltonian, and $\hat{F}_{\alpha }$ are the
matrices representing the Cartesian components of the spin angular momentum
with quantization axis fixed in the $z$-axis. The coupling constants $g_{n}$
and $g_{s}$ characterizing the density-density and spin-exchange
interactions, respectively, are related to the $s$-wave scattering lengths $%
a_{0}$ and $a_{2}$ in the total spin channels $F_{total}=0$ and $F_{total}=2$
by $g_{n}=4\pi \hbar ^{2}\left( a_{0}+2a_{2}\right) /3m$, $g_{s}=4\pi \hbar
^{2}\left( a_{2}-a_{0}\right) /3m$ \cite{Ho,Ohmi}. The ground state of the
spinor BEC depends crucially on the sign of $g_{s}$. For the \emph{%
ferromagnetic }coupling, $g_{s}<0$, the condensate is in the
\textquotedblleft axial\textquotedblright\ state, $\left\vert \left\langle
\mathbf{\hat{F}}\right\rangle \right\vert =1$ \cite{Ho}. On the other hand,
for the \emph{antiferromagnetic }coupling, $g_{s}>0$, the condensate is in
the \textquotedblleft polar\textquotedblright\ state, $\left\langle \mathbf{%
\hat{F}}\right\rangle =0$ \cite{Ho}. As we shall focus on the vortices
formed by the spin textures, it is more convenient to introduce the basis
set $\Psi _{\alpha }$ $(\alpha =x,y,z)$, such that $\Psi _{\pm 1}=\left( \pm
\Psi _{x}+i\Psi _{y}\right) /\sqrt{2}$ and $\Psi _{0}=\Psi _{z}$. As a
result, $\hat{F}_{\alpha }\left\vert \alpha \right\rangle =0$, and the spin
texture, which is parallel to the local magnetic moment, is defined by $%
\mathbf{S}\left( \mathbf{r}\right) =i\rho ^{-1}\mathbf{\tilde{\Psi}}%
^{\dagger }\times \mathbf{\tilde{\Psi}}$ where $\mathbf{\tilde{\Psi}=}\left(
\Psi _{x},\Psi _{y},\Psi _{z}\right) ^{T}$ \cite{Marchida}. Consequently, we
have $S_{x}\propto \textit{Re}\left[ \left( \Psi _{1}+\Psi _{-1}\right) \Psi
_{0}^{\ast }\right] $, $S_{y}\propto \textit{Re}\left[ i\left( \Psi _{1}-\Psi
_{-1}\right) \Psi _{0}^{\ast }\right] $, and $S_{z}\propto \left\vert \Psi
_{1}\right\vert ^{2}-\left\vert \Psi _{-1}\right\vert ^{2}$. For later
convenience, we define the unit vector $\mathbf{s}\left( \mathbf{r}\right) =%
\mathbf{S}\left( \mathbf{r}\right) /\left\vert \mathbf{S}\left( \mathbf{r}%
\right) \right\vert $.

\begin{figure}[htbp]\begin{center}
\includegraphics[width=3in]{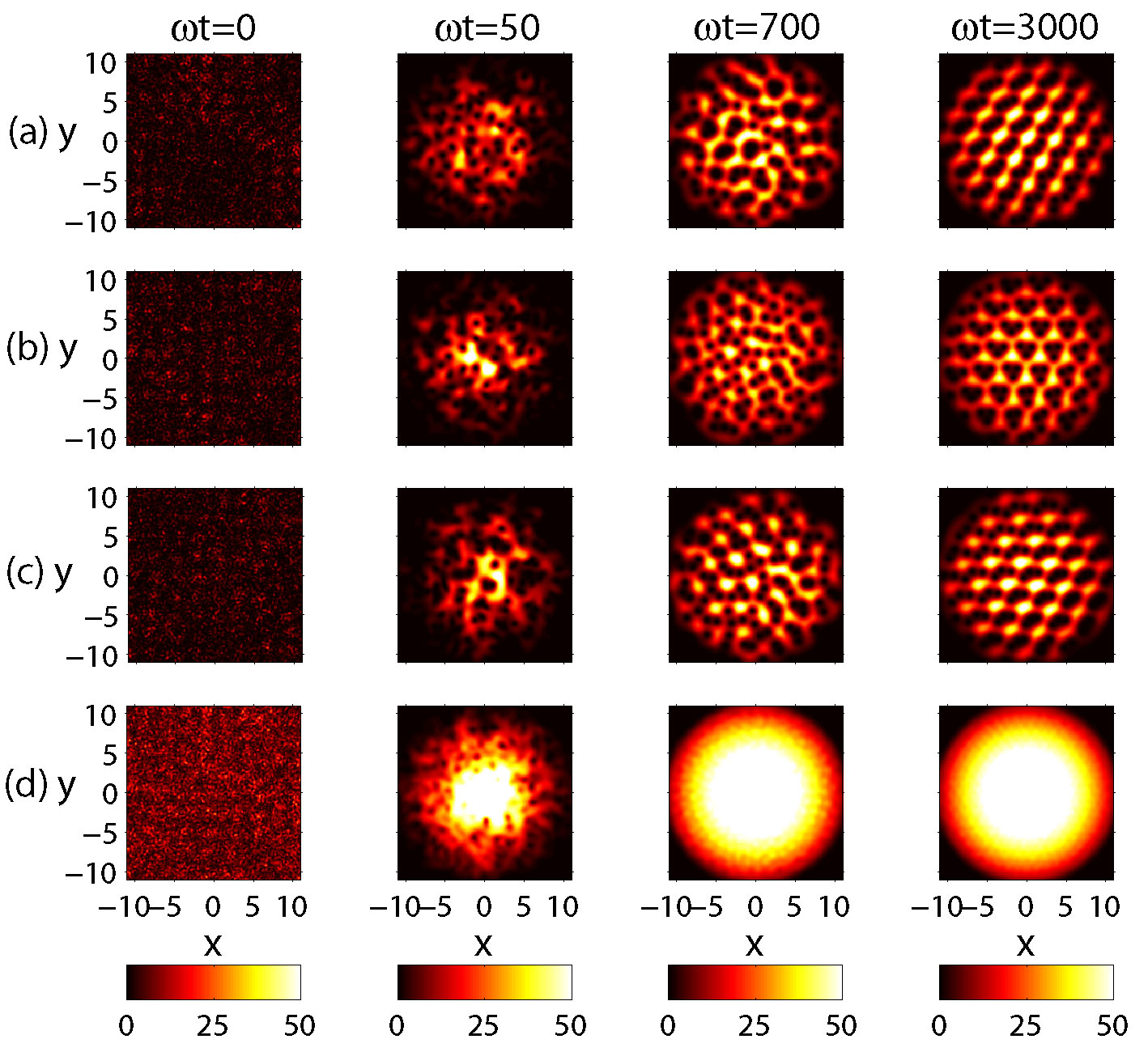}
\caption{(Color online) Snapshots of (a) $\left\vert \Psi _{-1}\right\vert ^{2}$; (b) $%
\left\vert \Psi _{0}\right\vert ^{2}$; (c) $\left\vert \Psi _{1}\right\vert
^{2}$; (d) $\left\vert \mathbf{\Psi }\right\vert ^{2}$ for the quenched
rotating spinor BEC of $^{87}$Rb. When the system reaches equilibrium at $%
T=10$ nK with $\Omega =0.95$, $\protect\mu =8$ (rightmost column),
crystalline order of vortex-trimers is established in each $\Psi _{j}$. The
particle numbers in the spinor BEC are $N_{\pm 1}\approx 4.84\times 10^{3}$,
$N_{0}\approx 4.79\times 10^{3}$.}
\label{sag}
\end{center}\end{figure}

\begin{figure}[htbp]\begin{center}
\includegraphics[width=3in]{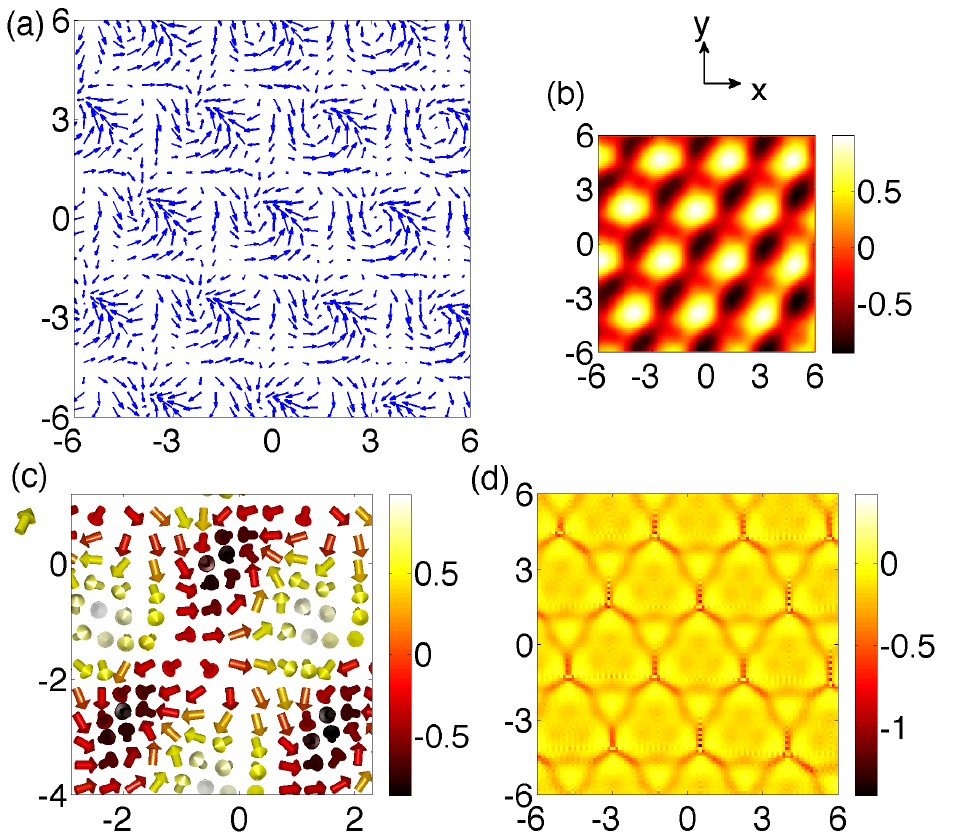}
\caption{(Color online) (a) The equilibrium
spin textures for the spinor BEC of $^{87}$Rb; (b) Distribution of $S_{z}$
in the spin textures; (c) The orientations of the unit vector $\mathbf{s}%
\left( \mathbf{r}\right) $ for a skyrmion. The color of each arrow indicates
the magnitude of $S_{z}$; (d) The density profile of the topological
charges.}
\label{sag}
\end{center}\end{figure}

To study the non-equilibrium dynamics of a quenched rotating spinor BEC, we
shall generalize the formulae in Ref.\cite{A. S. Bradley} to the following
set of coupled SPGPEs
\begin{eqnarray}
&&\partial _{t}\Psi _{j}  \label{SGPE} \\
&=&\mathcal{P}\{-\frac{i}{\hbar }\hat{H}_{j}^{GP}\Psi _{j}\mathbf{+}\frac{%
\gamma _{j}}{k_{B}T}\left( \mu -\hat{H}_{j}^{GP}\right) \Psi _{j}\mathbf{+}%
\frac{dW_{j}}{dt}\}  \nonumber
\end{eqnarray}%
where $T$ and $\mu $ denote the final temperature and chemical potential, $%
\gamma _{j}$ the growth rate for the $j$-th component, and $dW_{j}/dt$ is
the complex-valued white noise\ associated with the condensate growth. The
projection operator $\mathcal{P}$ restricts the dynamics of the spinor BEC
in the lower energy region below the cutoff energy $E_{R}$. In the rotating
frame, $\mathcal{\hat{H}}$ is replaced by $\mathcal{\hat{H}}-\Omega \hat{L}%
_{z}$, where $\hat{L}_{z}=-i\hbar \left( x\partial _{y}-y\partial
_{x}\right) $ is the $z$-component of the orbital angular momentum, and $%
\Omega $ is the angular frequency of rotation. As we shall focus on the fast
rotating BECs, in which the atomic cloud appears highly pancake-shaped, it
is reasonable to treat the system as two-dimensional. We therefore assume $V(%
\mathbf{r})=m\omega ^{2}\left( x^{2}+y^{2}+\lambda ^{2}z^{2}\right) /2$%
\textbf{\ }with $\lambda \gg 1$\textbf{. }The effective 2D interaction
strength can be obtained by integrating the wave functions with respect to $%
z $ to eliminate the axial degree of freedom\textbf{.} The numerical
procedures for integrating the set of coupled SPGPEs, are described as
follows. First, the initial state of each $\Psi _{j}$ is sampled by using
the grand-canonical ensemble for free ideal Bose gas at a temperature $T_{0}$
below the critical temperature and of chemical potentials $\mu _{j,0}$. The
spatial dependence of the initial state can be specified as a linear
combination of some basis functions. Here, we adopt the basis consisted of
plane waves with quantized momentum $\mathbf{k}=2\pi \left(
n_{x},n_{y}\right) /L$ ($n_{x}$, $n_{y}$ are integers and $L$ is the size of
the computation domain), i.e., $\Psi _{j}\left( t=0\right) =\sum_{\mathbf{k}%
}^{E_{R}}a_{j;\mathbf{k}}\psi _{\mathbf{k}}\left( \mathbf{r}\right) $, where
$\psi _{\mathbf{k}}\left( \mathbf{r}\right) $ are the plane-wave basis
functions. The condensate band lies below the an energy cutoff $E_{R}>E_{%
\mathbf{k}}=\hbar ^{2}\left\vert \mathbf{k}\right\vert ^{2}/2m$.
Furthermore, the distribution is sampled by $a_{j,\mathbf{k}}=\left( N_{j,%
\mathbf{k}}+1/2\right) ^{1/2}\eta _{j,\mathbf{k}}$ where $N_{j,\mathbf{k}}=%
\left[ \exp (\left( E_{j,\mathbf{k}}-\mu _{j.0}\right) /k_{B}T_{0})-1\right]
^{-1}$ and $\eta _{j,\mathbf{k}}$ are the complex Gaussian random variables
with moments $\left\langle \eta _{j,\mathbf{k}}\eta _{j,\mathbf{k}^{\prime
}}\right\rangle =\left\langle \eta _{j,\mathbf{k}}^{\ast }\eta _{j,\mathbf{k}%
^{\prime }}^{\ast }\right\rangle =0$ and $\left\langle \eta _{j,\mathbf{k}%
}\eta _{j,\mathbf{k}^{\prime }}^{\ast }\right\rangle =\delta _{\mathbf{kk}%
^{\prime }}$. Second, to simulate the thermal quench, the temperature and
chemical potential of the non-condensate band are altered to the new values $%
T<T_{0}$ and $\mu >\mu _{j,0}$. For convenience, we adopt the oscillator
units in the numerical computations, where the length, time and energy are
respectively scaled in units of $\sqrt{\hbar /m\omega }$, $\omega ^{-1}$ and
$\hbar \omega $.

We first study the spinor BEC of $^{87}$Rb, which has a $g_{s}<0$. The total
number of the modes are $n_{x},n_{y}=256$ and the energy cutoff is chosen at
$n_{xc},n_{yc}=128$. The parameters are $\Omega =0.95$, $T=10$nK, $\mu =8$,
and $\hbar \gamma _{j}/k_{B}T=0.03$ for all $\Psi _{j}$'s. The time
evolutions of the density profiles for $\Psi _{j}$'s are shown in Fig.1.
During the evaporative cooling, the rotating condensates grow up and the
emergent vortices start closely binding up and forming vortex-trimers in
each $\Psi _{j}$. When the system reaches equilibrium, these vortex-trimers
arrange themselves into some interwoven lattice structures such that each
vortex core of $\Psi _{j}$ is filled up with particles of the rest two
components. In other words, quantized vortices of either inter- or
intraspecies, avoid to overlap with each other. The shapes of the trimer
structures in all components are somewhat different. To characterize these
structures, we calculate the incompressible kinetic energy per particle for
each $\Psi _{j}$. This can be done by writing $\Psi _{j}=\left\vert \Psi
_{j}\right\vert \exp \left( i\varphi _{j}\right) $ and defining the current $%
\mathbf{Z}_{j}=\left\vert \Psi _{j}\right\vert \nabla \varphi _{j}=\mathbf{Z}%
_{j}^{\left( i\right) }+\mathbf{Z}_{j}^{\left( c\right) }$ \cite{Horng} in
terms of the solenoidal and irrotational fields, where $\nabla \cdot \mathbf{%
Z}_{j}^{\left( i\right) }=0$, $\nabla \times \mathbf{Z}_{j}^{\left( c\right)
}=0$. The incompressible energy is defined by $\mathcal{E}_{k,j}^{\left(
i\right) }=\left( 1/2\right) \int d^{2}r\left\vert \mathbf{Z}_{j}^{\left(
i\right) }\right\vert ^{2}$, which corresponds to the kinetic energy of
swirls in the superflows. Consequently, we find that $\mathcal{E}%
_{k,j}^{\left( i\right) }/N_{j}=19.48,$ $19.27,$ $20.05$ for $j=\pm 1,0$,
respectively. The spin textures are shown in Fig.2(a), where a hexagonal
lattice is visualized. Furthermore, the spatial variation of the local spin
moments parallel to the axis of rotation is plotted in Fig.2(b). An enlarged
perspective view of the 3-dimensional orientations of the local spins is
shown in Fig.2(c). Taking the island around the trap center for example, we
see that the innermost spin points into the paper, while the others
increasingly twist and bend in upwards direction. This vortex-like
arrangement of magnetic moments is exactly the configuration of a skyrmion
\cite{Yu,Muhlbauer,RoBler}. Since the central spin in the skyrmion is
perpendicular to the rotating plane, $S_{x}$ and $S_{y}$ must vanish
therein, implying that the skyrmions must be centered at the regions of $%
\Psi _{0}=0$, i.e., the cores of the vortex-trimer in $\Psi _{0}$. The
topological charge density, $\sigma =\mathbf{s\cdot }\left( \partial \mathbf{%
s/}\partial x\times \partial \mathbf{s/}\partial y\right) /4\pi $ \cite%
{Muhlbauer}, is shown in Fig.2 (d), which exhibits a hexagonal lattice
structure. Numerical integration over the primitive unit cell reveals that
each skyrmion carries a topological charge $Q=-1$. We notice that the
crystallization of skyrmions have recently been observed in various magnetic
materials characterized by the Dzyaloshinskii-Moriya interaction \cite%
{Muhlbauer,Yu}. Amazingly, the spin textures in these magnetic materials are
highly resembling to those in Fig.2(a).

\begin{figure}[htbp]\begin{center}
\includegraphics[width=3in]{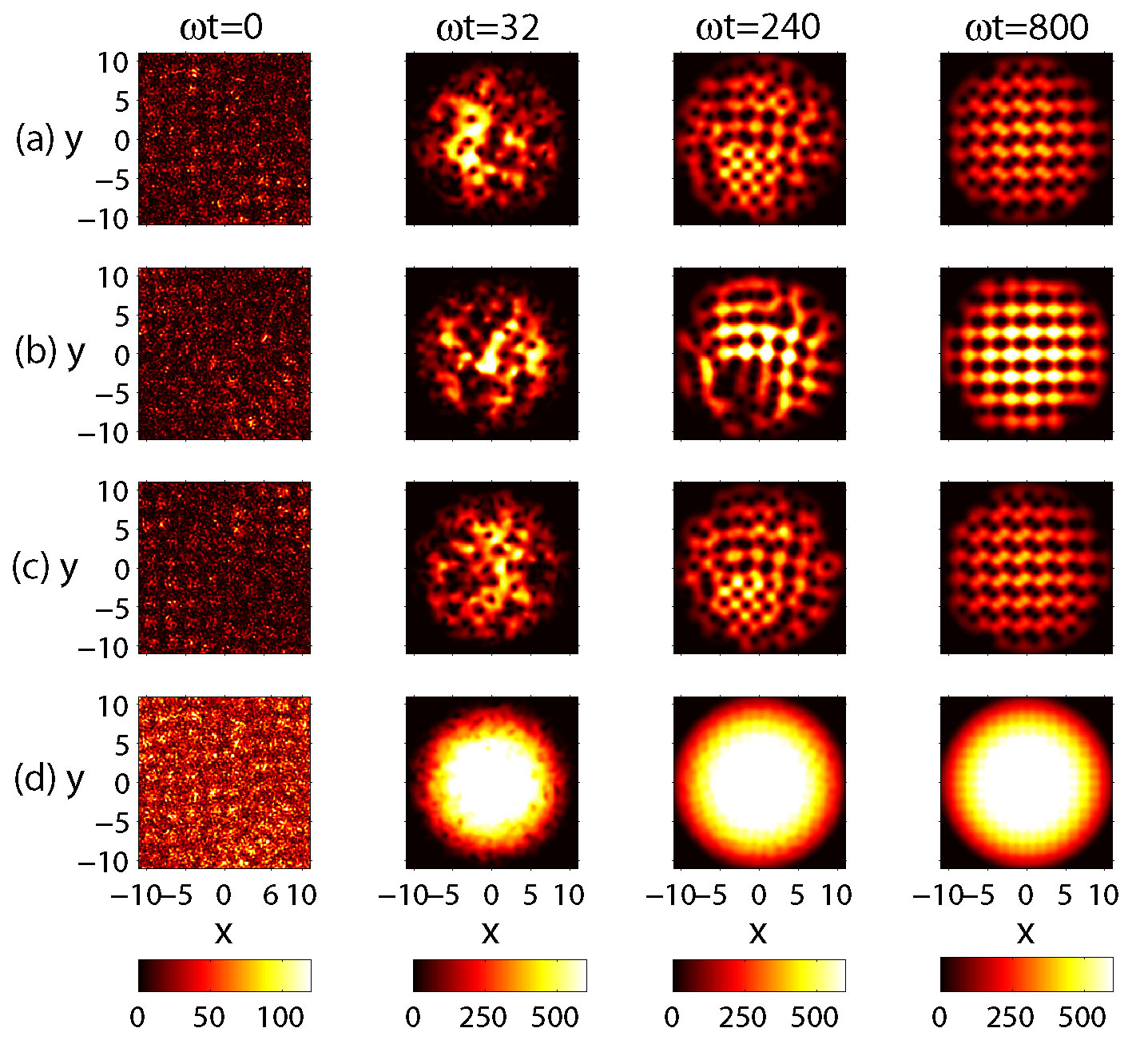}
\caption{(Color online) Snapshots of (a) $%
\left\vert \Psi _{-1}\right\vert ^{2}$; (b) $\left\vert \Psi _{0}\right\vert
^{2}$; (c) $\left\vert \Psi _{1}\right\vert ^{2}$; (d) $\left\vert \mathbf{%
\Psi }\right\vert ^{2}$ for a quenched rotating spinor BEC of $^{23}$Na.
When the system reaches equilibrium at $T=10$ nK with $\Omega =0.8$, $%
\protect\mu =25$ (rightmost column), crystalline order of vortex-dimers is
established in each $\Psi _{j}$. The particle numbers in the spinor BEC are $%
N_{\pm 1}\approx 4.55\times 10^{4}$, $N_{0}\approx 7.70\times 10^{4}$.}
\label{sag}
\end{center}\end{figure}

\begin{figure}[htbp]\begin{center}
\includegraphics[width=3in]{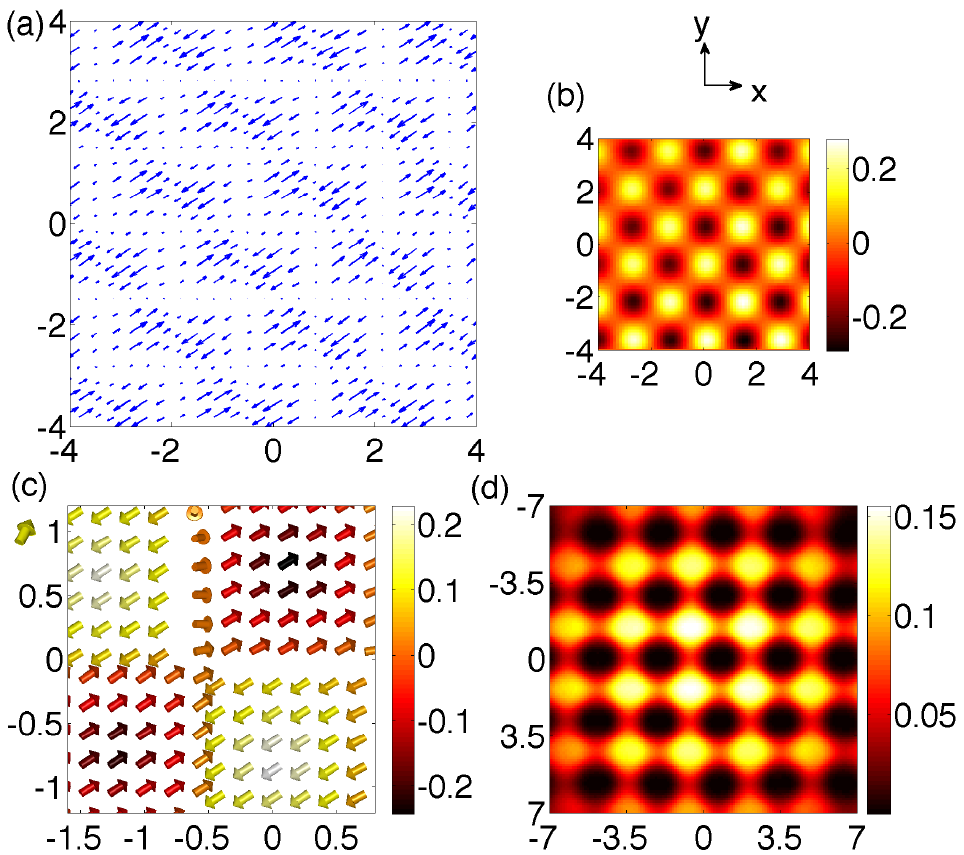}
\caption{(Color online) (a) The
equilibrium spin textures for the spinor BEC of $^{23}$Na; (b) Distribution
of $S_{z}$ in the spin textures; (c) The orientations of the unit vecotr $%
\mathbf{s}\left( \mathbf{r}\right) $ in the adjacent magnetic domains. The
color of each arrow indicates the magnitude of $S_{z}$; (d) The modulus of $%
\left\vert \Psi _{1}-e^{-2i\protect\phi \left( \mathbf{r}\right) }\Psi
_{-1}\right\vert $, where dark shaded regions reveal the locations of HQVs.}
\label{sag}
\end{center}\end{figure}

For the case of $g_{s}>0$, we consider the spinor BECs of $^{23}$Na. We set $%
\Omega =0.8$, $T=10$ nK, $\mu =25$, and $\hbar \gamma _{j}/k_{B}T=0.03$ for
all spin components. The total number of modes and energy cutoff remain the
same as those for the case of $^{87}$Rb. In Fig.3, the time evolutions of
the density profiles for $\Psi _{j}$ are shown. The nucleation of the
vortices in the current case are similar to that of the case of $^{87}$Rb,
except that a square lattice of tightly bound vortex-dimers is formed in
each component. Consequently, we find that $\mathcal{E}_{k,j}^{\left(
i\right) }/N_{j}=13.44$, $13.45$, $12.54$ for $j=\pm 1,0$, respectively. The
equilibrated spin textures on the rotating plan are shown in Fig.4(a), in
which a mosaic of magnetic domains with staggered magnetization is created.
In Fig.4(b) and (c), we see that almost all spins in the magnetic domains
lie in the $xy$-plane. Spins belonging to the same domain align nearly
unidirectionally. Note that the spins reverse their magnetization through a
Bloch wall transition in a very narrow region near the boundary between two
adjacent domains. These staggered magnetic domains act as the smoking gun of
HQVs which have been predicted to exist in superfluid $^{3}$He \cite{Volovik}
and superconductors \cite{Ivanov}. Considering the transformation, $\mathbf{%
\Psi }\rightarrow \hat{G}\left( \theta \right) \hat{R}\left( \mathbf{n},\chi
\right) \mathbf{\Psi }$, where $\hat{G}\left( \theta \right) =\exp \left(
i\theta \right) $ is a gauge transformation and $\hat{R}\left( \mathbf{n}%
,\chi \right) =\exp \left( i\chi \mathbf{n\cdot \hat{F}}\right) $ is a spin
rotation through an angle $\chi $ about the unit vector $\mathbf{n}$, a HQV
entails a spin rotation with $\chi =\pi $ followed by a global phase change
of $\pi $ in $\mathbf{\Psi }$. Without loss of generality, we assume $%
s\left( \mathbf{r}\right) =\cos \phi \left( \mathbf{r}\right) \mathbf{e}%
_{x}+\sin \phi \left( \mathbf{r}\right) \mathbf{e}_{y}$. Numerically, we
verify that the spin textures remain unchanged through the transformation $%
\hat{G}\left( \pi \right) \hat{R}\left( \mathbf{s},\pi \right) \mathbf{\Psi }%
=\left( e^{-2i\phi \left( \mathbf{r}\right) }\Psi _{-1},\Psi _{0},e^{2i\phi
\left( \mathbf{r}\right) }\Psi _{1}\right) ^{T}$. Upon requiring $\hat{G}%
\left( \pi \right) \hat{R}\left( \mathbf{s},\pi \right) \mathbf{\Psi =}%
e^{i2\pi }\mathbf{\Psi }$, it follows that $e^{-2i\phi \left( \mathbf{r}%
\right) }\Psi _{-1}=\Psi _{1}$ and $e^{2i\phi \left( \mathbf{r}\right) }\Psi
_{1}=\Psi _{-1}$. The redundancy of the last equality implies that there are
multiple solutions satisfying the criterion, $\left\vert \Psi
_{1}-e^{-2i\phi \left( \mathbf{r}\right) }\Psi _{-1}\right\vert =0$, or
equivalently, $\left\vert \Psi _{-1}\right\vert =\left\vert \Psi
_{1}\right\vert $. In Fig.4(d), the modulus $\left\vert \Psi _{1}-e^{-2i\phi
\left( \mathbf{r}\right) }\Psi _{-1}\right\vert $ is plotted where the dark
shaded areas represent the core positions of HQVs, which form a square
lattice apparently. Likewise, the HQVs are localized at the cores of
vortex-dimers in $\Psi _{0}$. Our results are consistent with those obtained
by means of dynamical creation \cite{Ji}, where a lattice of HQVs can be
created in a rotating optical trap when additional pulsed magnetic trapping
potentials are applied. Furthermore, the ground state of a rotating dipolar
spinor BEC has been shown to have the same structure when the dipole-dipole
interaction is small compared to the contact ones \cite{Simula}. \

We note that an $\Omega $ comparable to $\omega $ is needed to stabilize the
crystalline orders of defects. Under such a fast rotation, the filling
factor $\nu $, i.e., the ratio of the number of atoms to the number of
vortices, can have a value of few hundreds for each component, as shown in
Fig.1 and Fig.3. According to the criterion in \cite{Hall}, the system
enters the mean-field quantum Hall regime, in which the mean-field theory
still applies yet the state of the system can be well described in the
lowest Landau level approximation. When $\Omega $ is not sufficiently large,
the crystallization does not arise albeit some few topological defects may
be readily created in the condensate. For example, in our simulations with $%
\Omega =0.3$, we find only a few HQVs nucleating in the spinor BEC of $^{23}$%
Na during the rapid rotational evaporative cooling. The situation is
somewhat different in the case of $^{87}$Rb, where Mermin-Ho vortices,
rather than the skyrmions, are created in the condensate. Furthermore, to
see whether the crystallization is robust against the thermal fluctuations
arising from the growth of condensate, we have assumed a range of final
equilibrium temperatures in the simulations. We find that both crystalline
orders remain intact for temperatures lower than $50$nK. At higher
temperatures, however, the crystallization is thwarted by the fluctuations
of spin textures so that the lattice becomes disordered and starts melting
when the system approaches the critical regime.

In summary, we have investigated the non-equilibrium dynamics of spin-1 BECs
during the rapid rotational evaporative cooling. Crystallization of
skyrmions and HQVs is predicted to arise in the spinor BEC of $^{87}$Rb and $%
^{23}$Na, respectively. To resolve the spatial magnetization of the
crystallized topological defects, the images have to be taken in situ, and
this can be achieved basically by using the polarization-dependent
phase-contrast technique \cite{Higbie}.

S.-C. Gou is supported by NSC, Taiwan, under Grant No.
98-2112-M-018-001-MY2. W. M. Liu is supported by NSFC under Grant No.
10934010, and NKBRSFC under Grant No. 2011CB921502.

\end{document}